\documentclass[10pt,letterpaper,twocolumn]{article} 

\usepackage{ol2}
\usepackage[colorlinks=true,urlcolor=blue, linkcolor=blue]{hyperref}
\usepackage{amsmath}

\begin{document}
\twocolumn[
 
\title{Broadband amplitude squeezing in a periodically poled $\mathbf{KTiOPO_4}$ waveguide}
\author{Matthew Pysher, Russell Bloomer, and Olivier Pfister$^{*}$}
\address{Department of Physics, University of Virginia, 382 McCormick Road, Charlottesville, VA 22904-4714, USA}
\author{Christopher M. Kaleva, Tony D. Roberts, and Philip Battle}
\address{AdvR Inc., 2310 University Way, Bldg.~\#1, 
 Bozeman, MT 59715, USA}
\address{$^*$Corresponding author: opfister@virginia.edu}

\begin{abstract}
We generated -2.2 dB of broadband amplitude squeezing at 1064 nm in a periodically poled $\rm KTiOPO_4$ (PPKTP) waveguide, by coupling of the fundamental and second harmonic continuous-wave fields. This is the largest amount of squeezing obtained to date in a KTP waveguide, limited by propagation losses. This result paves the way for further improvements by use of lower-loss buried ion-exchanged waveguides.
\end{abstract}
\ocis{270.6570, 270.1670, 190.7220}
\maketitle
]
The experimental implementation of continuous-variable (CV) quantum information \cite{Braunstein2005}, an ambitious and exciting endeavor, requires the creation of strongly squeezed light. 
Squeezed light has been produced using a number of methods, but the most successful experiments to date (ranging from -9 to -10 dB of squeezing) have used optical parametric oscillators (OPO), which feature a second-order nonlinear material placed in a resonant optical cavity \cite{10db,laurat,9db}. In such systems, intracavity losses are amplified by the resonator buildup and thus present a serious hindrance to increasing the squeezing level. It would therefore be beneficial to suppress the optical cavity by use of nonlinear optical waveguides, in which the transverse field confinement yields an increase of the nonlinear efficiency that can make up for the buildup of a reasonably high finesse cavity \cite{param, advr}. If need be, some cavity modes can still be exquisitely well defined by seeding the nonlinear waveguide with an optical frequency comb \cite{Hall2006,Hansch2006}. Waveguides are ideally suited for applications such as integrated circuits, due to their small size \cite{kanter}, and  could also help alleviate gain-induced diffraction which has been seen with traveling waves in bulk crystals \cite{kim}. Moreover, removing the optical cavity yields an increase of the squeezing bandwidth by several orders of magnitude \cite{furusawa}, which is of interest for fast quantum processing. Finally, comb-seeded waveguides are of great interest for a recently proposed method to generate massively scalable CV entanglement \cite{Menicucci2008}.

Over a decade ago, several experiments tried to exploit the increase in nonlinear efficiency that waveguides provide in an attempt to obtain large amounts of traveling-wave squeezing with pulsed inputs \cite{anderson95, anderson97, serkland95, serkland97}.  However, due to propagation losses in the waveguide, these works achieved a maximum of -1.5 dB of squeezing.  Recently, advances in waveguide fabrication techniques have allowed for better than -4 dB of pulsed traveling-wave squeezing in MgO-doped periodically poled $\rm LiNbO_3$ (PPLN) waveguides \cite{eto}.  Undoped PPLN waveguides were used to obtain squeezing and entanglement with a continuous-wave (CW) input \cite{furusawa}.  Other media for nonlinear optical waveguides include quasi-phase-matched KTP \cite{anderson95}, quasi-phase-matched $\rm LiTaO_3$ (LT) \cite{anderson97}, and periodically poled stoichiometric LT \cite{ppslt}.  Although the first measurement of squeezed light from an optical waveguide was made in KTP \cite{anderson95}, squeezing in KTP waveguides has not been explored in recent years.  There are, however, a number of reasons to do so: high-squeezing experiments have been carried out in bulk KTP and PPKTP \cite{9db, laurat}, KTP has a high damage threshold, a low amount of green induced IR absorption \cite{wang, batchko}, low temperature sensitivity \cite{bierleinvan, bierleinarw}, and no significant photorefraction damage \cite{eger}. All this makes KTP a very promising material for nonlinear optical waveguide applications.  	

In this letter, we report -2.2 dB of amplitude squeezing at 1064 nm from a PPKTP waveguide.  This is the largest amount of squeezing obtained from a PPKTP waveguide, to the best of our knowledge, and it is also the largest amount of squeezing reported from traveling-CW optical parametric amplification. 

The 9.5 mm long PPKTP waveguides are manufactured by Advanced Research (AdvR), Inc.  They are fabricated using photolithography on a flux grown, $Z$-cut KTP wafer. A direct contact mask designed for 4 $\mu$m wide channel waveguides is used to pattern a layer of Al onto the +$Z$ surface of the wafer.  Patterning KTP waveguides was done using the nanofabrication user facilities at the University of California-Santa Barbara (UCSB) which is part of the National Nanofabrication Infrastructure Network (NNIN).  The wafer is diced into 2 mm $\times$ 10 mm chips and polished on the optical surfaces. The chips are placed in a molten bath of $\mathrm{RbNO_3}$ salt at 400$^\circ$C for 120 minutes. The bare areas of the patterned surface undergo ion exchange in which Rb$^+$ ions diffuse into the KTP, replacing K$^+$ ions to a depth of 4-9 $\mu$m, forming a surface index step of approximately  0.02 relative to the surrounding KTP.
	
Periodic electrodes were fabricated on a separate quartz substrate using contact lithography to define a chrome grating pattern with an 8.09 $\mu$m period.  The patterned electrode was aligned and pressed to the +$Z$ surface; a ground electrode consisting of a uniform metal substrate contacted the -$Z$ surface.  The poling waveform was applied using a Trek 20/20C high voltage amplifier controlled by a computer program that simultaneously monitored the electrode current. 

Our experimental setup is shown in Fig.1.  Both 532 and 1064 nm fields were generated by an Innolight ``Diabolo'' CW Neodymium-doped Yttrium Aluminum Garnet laser with an external resonant frequency doubler.  The 532 and 1064 nm beams were combined on a dielectric mirror and coupled into the waveguide by a 10$\times$ microscope objective of effective focal length of 1.537 cm.  A telescope was placed in the path of the green beam to correct the focal length change of the microscope objective at 532 nm due to dispersion, so that we could place the focus of both beams precisely at the input facet of the waveguide chip.  The temperature of the waveguide was controlled to the nearest hundredth of a degree by a temperature lock loop (Wavelength Electronics LFI-3751).  The fundamental light exiting the waveguide was collimated using an aspherical lens, of focal length 8 mm, before being sent to a balanced homodyne detection system consisting of a half-wave plate, a polarizing beam splitter, and two low-noise photodetectors (JDSU ETX 500 InGaAs PIN photodiodes, quantum efficiency of about 94\% at 1064 nm).  The preamplified signals from the detectors were then added and subtracted to yield the amplitude-squeezed and shot noise levels, respectively \cite{yuen}.

\begin{figure}[htb]
\begin{center}
\includegraphics[width=3.25in]{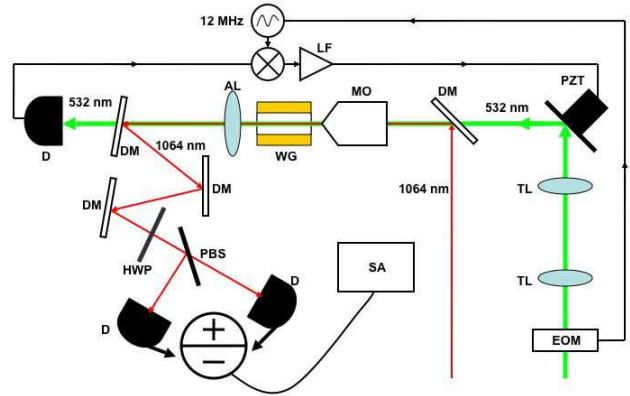}
\end{center}
\caption{(Color online) Experimental squeezing setup with AL, aspheric lens; D, detector; DM, dielectric mirror; HWP, half-wave plate; LF, loop filter; MO, microscope objective; PBS, polarizing beam splitter; PZT, piezoelectric transducer; SA, spectrum analyzer; WG, waveguide.}
\label{setup}
\end{figure} 

Whether up- or downconversion takes place between the two optical frequencies inside the nonlinear crystal is decided by the compound phase $\phi_{2\omega}-2\phi_{\omega}=\pm\pi/2$, respectively \cite{shen}. The interaction was servo controlled to either possibility by detecting the beat note of the incident green beam, frequency-modulated at 12 MHz by an electro-optic modulator (EOM), with the second harmonic beam created in the crystal. A dichroic mirror after the waveguide separated out the light at 532 nm which was detected by a Si photodiode. After demodulation, the error signal was sent to a proportional-integrator loop filter and applied to a piezoelectrically actuated mirror so as to control the relative optical path of the two fields.
 
The PPKTP waveguide's second-harmonic-generation efficiency was measured to be on the order of 1 W$^{-1}$cm$^{-2}$. Parametric amplification was measured by coupling 5 mW of IR and 335 mW of green light into the waveguide, of which 2.9 mW of IR and 195 mW of green were transmitted, i.e.\ a 58\% throughput for either wavelength. This figure should account for Fresnel losses on the uncoated crystal facets as well as propagation losses from waveguide imperfections. 
Both fields were $Z$-polarized in the crystal. We observed an amplification of the IR output power by a factor of 2.38 when the phase was set for parametric downconversion, and a deamplification by a factor $\rho_{cl}=\langle P_{\mathit{out}} \rangle/\langle P_{\mathit{in}}\rangle = 0.45$ when the phase was set for upconversion. No damage was observed in the crystal despite the large amount of CW green power. 

For moderate amplitude squeezing of a pure coherent state $|\alpha\gg 1\rangle$, $\langle N_{\mathit{out}}\rangle = \langle N_{\mathit{in}}\rangle \exp(-2r) + \sinh^2 r \simeq |\alpha|^2 \exp(-2r)$, where $r$ is the squeezing parameter. Using $\rho_{cl}=\exp(-2r)$, we would therefore expect to observe $10\log(\rho_{cl})=-3.5$ dB of squeezing. In a more realistic, if still simplified, analysis, we take into account the Fresnel transmissivity of the uncoated output facet $\eta_f=91\%$, the detectors' finite quantum efficiency $\eta_d=94\%$, the propagation efficency after the waveguide $\eta_p=98\%$, and the unknown waveguide propagation loss $\eta_w$ to express the measured squeezing factor of -2.2 dB in terms of the detected photon-number variances $\rho=V(N_2+N_3)/V(N_2-N_3) = \eta\rho_{\mathit{cl}} + 1-\eta$, where $\eta=\eta_w\eta_f\eta_d\eta_p$. This yields $\eta_w=86(2)\%$, i.e.\ an estimated $-0.6(2)$ dB/cm waveguide loss.

Squeezing was observed over 20 MHz, as displayed in Fig.2, the bandwidth being solely limited by our detection system. 
\begin{figure}[htb]
\begin{center}
\includegraphics[width=3.25in]{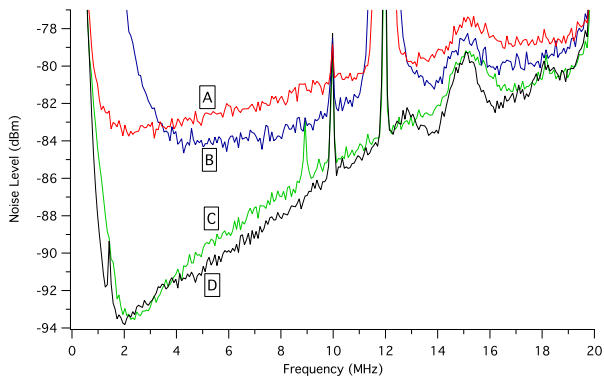}
\end{center}
\caption{(Color online) Squeezing data with A, shot noise; B, squeezing; C, electronics noise for shot noise measurement; and D, electronics noise for squeezing measurement.  The resolution and video bandwidths are 100 kHz.  All traces are averaged 300 times.}
\label{squeezing}
\end{figure} 
With a broader-band detection system, our squeezing bandwidth would be limited by the much larger phasematching bandwidth ($\sim 100$ GHz or more). The maximum amount of squeezing is observed between 7 and 9.5 MHz.  We observe -1.86 dB of raw squeezing over this range, i.e.\ -2.22 dB once the electronics noise is subtracted out.  Note that this squeezing would be readily increased to -2.7 dB just by adding antireflection coatings to the output face of the waveguide and photodiodes' surfaces (thereby setting $\eta_f=\eta_d=1$). However, the primary goal of this work was to assess the effect of the waveguide's propagation loss $\eta_w$ on the squeezing.  While this loss is already low, it is the dominant one and needs to be addressed. Future steps to decrease coupling and propagation losses will include burying the waveguides, adding tapers, and increasing patterning tolerances. These improvements are expected, in turn, to increase the squeezing level.   

In conclusion, we have observed -2.2 dB of amplitude squeezing in a PPKTP waveguide by seeded traveling-wave SHG with CW inputs.  The ability to obtain large amounts of squeezing from waveguides could allow for the elimination of optical cavities from many key experiments in the areas of quantum information and quantum computing \cite{Menicucci2008}. Because of its excellent optical properties and low bulk loss, KTP is a promising material for the development of ultralow loss nonlinear waveguides.

We would like to thank Martin Fejer and Mirko Lobino for useful discussions.  MP and OP were supported by National Science Foundation grants No. PHY-0555522 and No. CCF-0622100.  AdvR was supported in part by the U.S. Air Force Research Laboratory.

\end{document}